\documentclass[pra,letterpaper,preprint]{revtex4-1} 

\usepackage{graphicx}
\usepackage{dcolumn}
\usepackage{bm}
\usepackage{amsbsy,amssymb,amsmath,bm,bbold}
\usepackage{multirow}
\usepackage{graphicx,color,epsfig,rotate}
\usepackage{times,xspace}
\usepackage{fancyhdr}  
\usepackage{epstopdf}
\usepackage[justification=justified]{caption}

\begin{document}


\title{Nonadiabatic Modal Dynamics Around a Third-order Exceptional Point in a planar waveguide}%

\author{ Sibnath Dey$^1$, Arnab Laha$^{1,2}$, Somnath Ghosh$^1,$}
\email{somiit@rediffmail.com}
\affiliation{$^1$Unconventional Photonics Laboratory, Department of Physics, Indian Institute of Technology Jodhpur, Rajasthan 342037, India.\\ $^2$Institute of Radiophysics and Electronics, University of Calcutta, Kolkata 700009, India}


\begin{abstract}
Dynamical parametric encirclement around an Exceptional Point (EP) and corresponding asymmetric state transfer phenomenon have attracted considerable attention recently. In this context, beyond the reported time-asymmetric state dynamics around a second-order EP (EP2) in a two-level system, the investigation of similar state dynamics around a third-order EP (EP3) in a multi-state system, having comparably complex topology with rich physics, is lacking. Here, we report a fabrication-feasible few-mode planar optical waveguide with a customized gain-loss profile and investigate the effect of dynamical parametric encirclement around an EP3 in the presence of multiple EP2s. The cube-root branch point behavior is established in terms of successive switching between the propagation constants of the coupled modes following an adiabatic encirclement process. Now, while considering the dynamical encirclement process, the breakdown in system adiabaticity around an EP3 leads to a unique light dynamics, where we have shown the breakdown of chirality  of the device.
\end{abstract}

\maketitle

The unconventional aspects of non-Hermitian quantum mechanics are widely used to demonstrate open quantum-inspired photonic devices. In a non-Hermitian system, the eigenvalues are complex, and corresponding eigenstates form a non orthogonal set. One of the most interesting phenomena in non-Hermitian quantum mechanics is Exceptional Points (EPs) in the parameter space of a physical problems. At the EP, not only the eigenvalues but also eigenvectors coalesce ~\cite{Mirieaar7709,Ghosh16,isolator_our,Laha18}, unlike the diabolic points (DP) in standard Hermitian system, where only eigenvalues coalesce but the eigenvectors remain different. Unconventional properties of EPs have been observed in experiments ~\cite{PhysRevX.8.021066,PhysRevE.69.056216}. Unconventional behaviours of EPs have been extensively studied in optical waveguides~\cite{Ghosh16,Laha18,isolator_our}, micro cavities~\cite{Laha_2019,Laha:17}, lasers~\cite{doi:10.1063/1.5040036}, photonic crystals~\cite{PC_slabs}, and also in several non-optical systems like atomic~\cite{atomic_spectra} and molecular spectra~\cite{PhysRevLett.103.123003}, etc. With proper control over fabrication technique, EP in photonic systems offers itself as a strong competitor to meet the present-day challenges like unidirectional light propagation with enhanced non reciprocity~\cite{isolator_our,unidirectional_light}, asymmetric mode conversion~\cite{Ghosh16,isolator_our,Laha18}, ultra sensitive EP aided sensing~\cite{ep_sensors}, etc.  

 One of the consequences of the state coalescence is that the eigen energies presented in the parameter space exhibit a branch-point singularity at the EP. When two external parameters are varied to form a closed trajectory enclosing an EP in the parameter space, the two eigenvalues, exchange energies between each other~\cite{Mirieaar7709,Ghosh16,isolator_our,Laha18,PhysRevX.8.021066}. Here, a total 2$\pi$ rotation around an EP in the parameter space results in permutation between the corresponding states (i.e., interchange of initial positions). Thus, a complete 4$\pi$ rotation around an EP is needed to recover the original state except for geometric phase of  $\pm\pi$~ \cite{refId0}. This is topological because it occurs only if the parametric loop encloses the EP irrespective of its precise shape. In this context, successive state exchange phenomenon around a higher order EP has also been studied in the literature, even though mostly theoretical \cite{Demange_2011,Heiss_2008,EP3_sayan,EP4_sayan}. If we consider the length or time dependent parametric evolution to enclose an EP dynamically, then adiabaticity breaks down during encirclement~\cite{PhysRevA.88.010102}. Here an anticlockwise and a clockwise parametric evolution results in different dominating state at the  output~\cite{Ghosh16}. If we consider two interacting modes, the mode which encounter with lower average loss, evolves adiabatically. In this context, the state dynamics during the dynamical parametric evolution around a higher order EP in a single planar geometry is an important issue and need to be explored. Beyond the already reported three mode coupled waveguide system ~\cite{Zhang19}, it would be interesting and more compact from fabrication point-of-view in integrated devices for mode management in multi-mode planar geometry. Using minimum number of control parameters to locate an EP3 with associated successive state conversion in a simple fabrication feasible system is a challenge. 

In this latter, to address the issues mentioned above, we investigate a specially configured gain-loss assisted few-mode supported planar optical waveguide to host an EP3. Modulating the spatial distribution of a multilayer gain-loss profile via two tunable parameters, we have controlled the coupling between the quasi-guided modes, where we have encountered an EP3 in the simultaneous presence of two EP2s between three chosen interacting modes. The cube-root branch point behavior of the embedded EP3 has been verified with the successive switching between the propagation constants ($\beta$-values) of three interacting modes following an adiabatic parametric variation around two corresponding EP2s. We also investigate the propagations of three interacting modes around the EP3, following a dynamical parametric encirclement process. Here, we have shown the nonchiral dynamics of the interacting modes in the sense that a specific mode dominates at the end of the encirclement process, irrespective of the direction of encirclement. Beyond the coupled waveguide systems, the EP3-driven modal dynamics in our planar waveguide structure is indeed more suitable for device applications in integrated photonic circuits.

\begin{figure}[t]
\centering
\includegraphics[width=12cm]{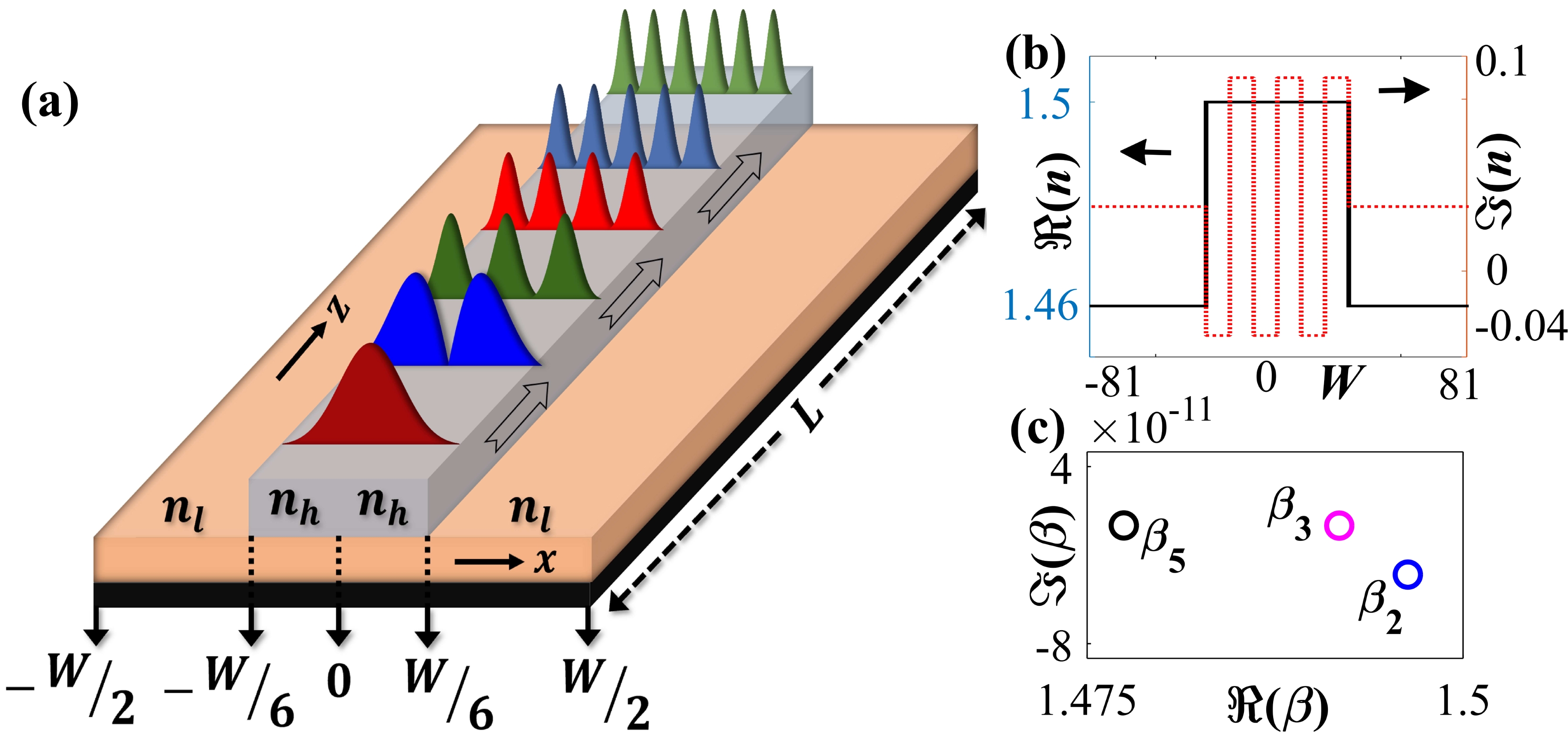}
\caption{\textbf{(a)} Schematic of the proposed waveguide. $x$- and $z$-axis represents transverse and propagation directions, respectively. \textbf{(b)} Transverse refractive index profile $\Re(n)$ (black line) corresponds to left vertical axis and $\Im(n)$ at specific $\gamma$=0.006 and $\tau$=3.8 (dotted red line) corresponds to right vertical axis. \textbf{(c)} Nonlinear initial distribution of propagation constants ($\beta$-values) of three chosen interacting modes $\beta_j\,(j=2,3,5)$ in complex $\beta$-plane.}
\label{fi1}
\end{figure}

We design a step index  planar optical waveguide, schematically shown in Fig. \ref{fi1}(a). Refractive indices of  cladding and core are chosen as $n_l=1.46$ and $n_h=1.50$, respectively. We normalize the operating frequency $\omega=1$ and set the width $W=162\lambda/2\pi=162$ and length $L=15\times 10^3$ in dimension less unit for which waveguide host six quasi-guided scalar modes, say $\psi_j\,(j=1,2,3,4,5,6)$. The waveguide occupies the region $-W/2\le x\le W/2$ along the transverse ($x$) direction and $0\le z\le L$ along the longitudinal ($z$) direction. To control the mutual coupling among the six quasi guided modes simultaneously, we introduce non-hermiticity in terms of a spatial multilayer unbalanced gain-loss profile in such a way that the overall complex refractive index profile can be written as:
\begin{equation}
n(x)=\left\{ 
\begin{array}{ll}\vspace{0.2cm}
n_l+i\gamma,\quad&\textnormal{for}\,\,\,W/6\le |x|\le W/2\\
n_h-i\gamma,\quad&\textnormal{for}\,\left\{ 
\begin{array}{l}-W/4\le x\le -W/6\\-W/8\le x\le 0\\W/8\le x\le W/6 \end{array}\right.\vspace{0.2cm}\\
n_h+i\tau\gamma,\quad&\textnormal{for}\,\left\{ 
\begin{array}{l} -W/6\le x\le -W/8\\0\le x\le W/8\\W/6\le x\le W/4 \end{array}\right.
\end{array}
\right.
\label{nx} 
\end{equation}
Here, $\gamma$ and $\tau$ represents two coupling control parameters, where $\gamma$ is gain-coefficient, and  $\tau$ is fractional loss-to-gain ratio. Thus, inside the core, ($-W/6\le x\le W/6$) we have introduced six alternative layers of gain and loss having equal widths, and inside the cladding ($W/6\le |x|\le W/2$) only loss. The overall refractive index distribution for a specific set of $\{\gamma,\tau\}$ has been shown in the upper panel of the Fig. \ref{fi1}(b). We can modulate system nonhermiticity by tuning  $\tau$ and $\gamma$ independently (maintaining the Kramers–Kronig with the consideration of single frequency operation) \cite{Laha18}. We compute propagation constants $\beta_j(j=1,2,3,4,5,6)$ of the supported modes using the scalar modal equation $[\partial_x^2+n^2(x)\omega^2-\beta^2]\psi(x)=0$ (with the approximation of small $\Delta n=n_h-n_l$). With the judicious choice of system parameters, nonlinear initial distribution of $\beta$-values of a set of three chosen interacting modes has been showing in complex $\beta$-plane in Fig. \ref{fi1}(c).
\begin{figure}[t]
\centering
\includegraphics[width=12cm]{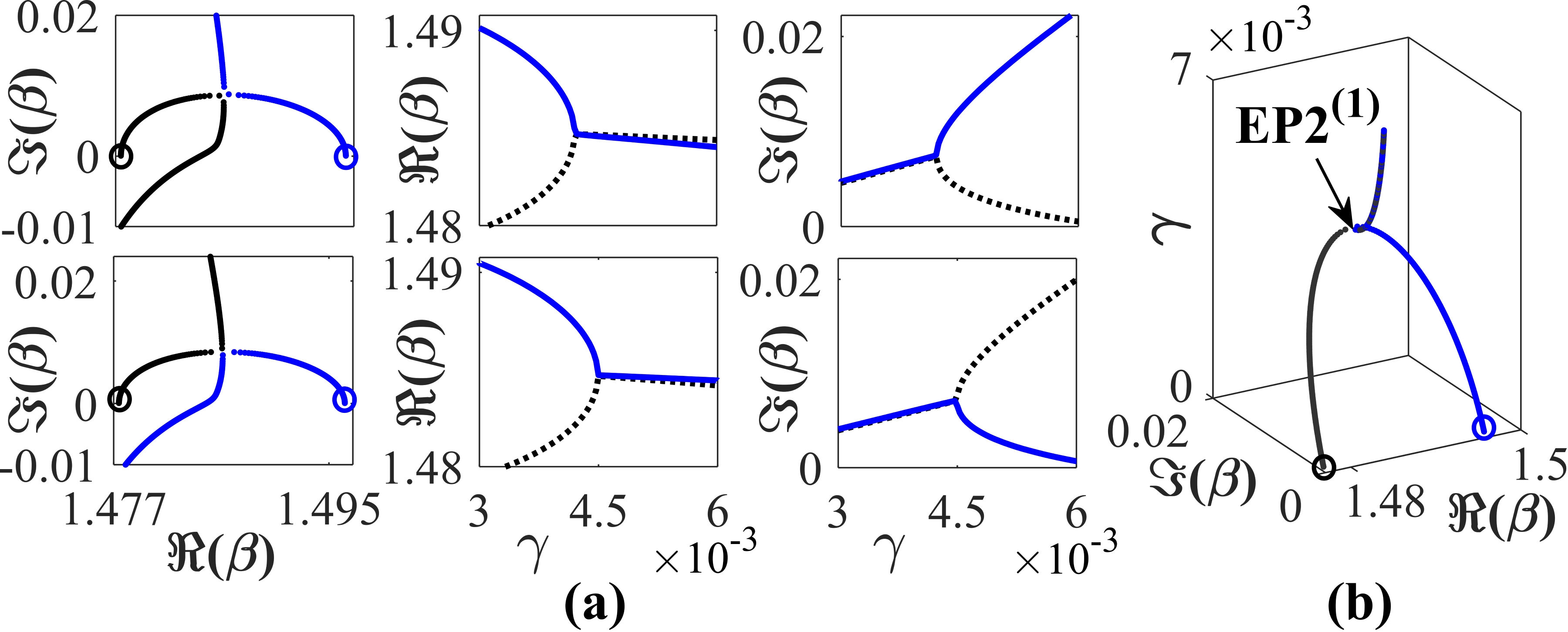}
\caption{\textbf{(a)} Dynamics of $\beta_2$ and $\beta_5$ (blue and black dotted line) exhibiting ARCs for different $\tau$-values with an increasing $\gamma$. Upper panel represents the ARC for $\tau$=5.2 with a crossing in $\Re(\beta)$ and an anticrossing in $\Im(\beta)$. Lower panel represents the ARC for $\tau$=4.8 with an anticrossing in $\Re(\beta)$ and a crossing in $\Im(\beta)$. \textbf{(b)} Coalescence of $\beta_2$ and $\beta_5$ near $\gamma\approx0.0043$ for $\tau=5$. Circles with respective colour denote the initial positions (i.e, for $\gamma=0$) of $\beta_2$ and $\beta_5$ }.
\label{fig2}
\end{figure}

Now, with introduction of gain-loss, all the supported modes are mutually coupled. Here, we host an EP3 with the encounter of two EP2s in ($\gamma,\tau$)-plane among three interacting modes. To encounter an EP2, we use the concept of special avoided resonance crossing (ARCs) \cite{Ghosh16} in complex $\beta$-plane between two interacting modes from the chosen set. In Fig. \ref{fig2}, we have shown such a special ARC phenomena between $\beta_2$ and $\beta_5$ for different $\tau$ values with in a range of $\gamma$ from 0 to 0.006. Here blue and black curves indicate the trajectories of $\beta_2$ and $\beta_5$, respectively. We identify an ARC at $\tau=5.2$, where, $\Re[\beta]$ undergoes crossing and $\Im[\beta]$ undergoes anticrossing, as shown in the plots in upper panel of Fig. \ref{fig2}(a). A different kind of ARC has been identified for $\tau=4.8$, where, $\Re[\beta]$ undergoes an anticrossing and $\Im[\beta]$ undergoes a crossing, as shown in the plots in lower panel of Fig. \ref{fig2}(a). This certain transition between two topologically dissimilar behaviour of ARCs for two different values of $\tau$ illustrated in upper and lower panel of Fig. \ref{fig2}(a), clearly indicates the appearance of an EP, where two coupled states are analytically connected \cite{Ghosh16}. We locate the approximate position of this branch point with judicious choice of an intermediate value of $\tau=5$ and track the dynamics of the complex $\beta$-values in Fig. \ref{fig2}(b). Here, we have shown that $\beta_2$ and $\beta_5$ coalesce at $\tau\approx0.0043$. Such a coalescence in complex $\beta$-plane certainly indicate the presence of a second order EP (say, EP2$^{(1)}$) in $(\gamma,\tau)$-plane at $\sim(0.0043,\,5)$ In a Similar way , we find out another location of EP2 in $(\gamma,\tau)$ plane at $\sim(0.012,\,2.9)$ (say, EP2$^{(2)}$) between $\beta_2$ and $\beta_3$. 

We investigate the effect of encirclement around a single EP2 or both with proper choice of closed parameter space. To enclose single or multiple EP2s, we choose a closed length dependent elliptical parametric variation in $(\gamma,\tau)$ plane following the equations
\begin{equation}
\gamma(\phi)=\gamma_{0}\sin\left(\phi/2\right);\,\,\tau(\phi)=\tau_{0}+a\sin\left(\phi\right).
\label{device} 
\end{equation}
\begin{figure}[t]
\centering
\includegraphics[width=12cm]{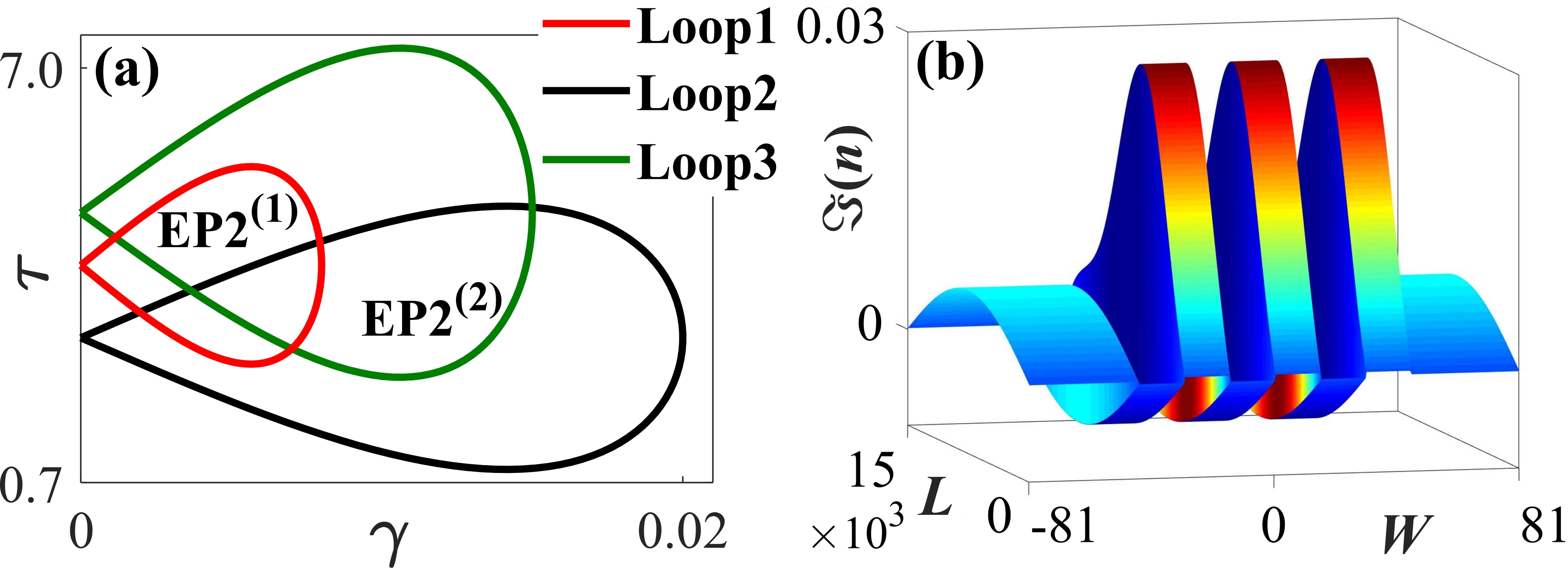}
\caption{\textbf{(a)} Three chosen parametric loops in ($\gamma$,$\tau$)-plane to encircle two EP2s individually (Loop1 and Loop2) and also simultaneously (Loop3). \textbf{(b)} Overall distribution of $(\Im[n(x,z)])$ after mapping Loop3 shown in (a) throughout the waveguide.}
\label{fig3}
\end{figure}
\begin{figure}[b!]
\centering
\includegraphics[width=12cm]{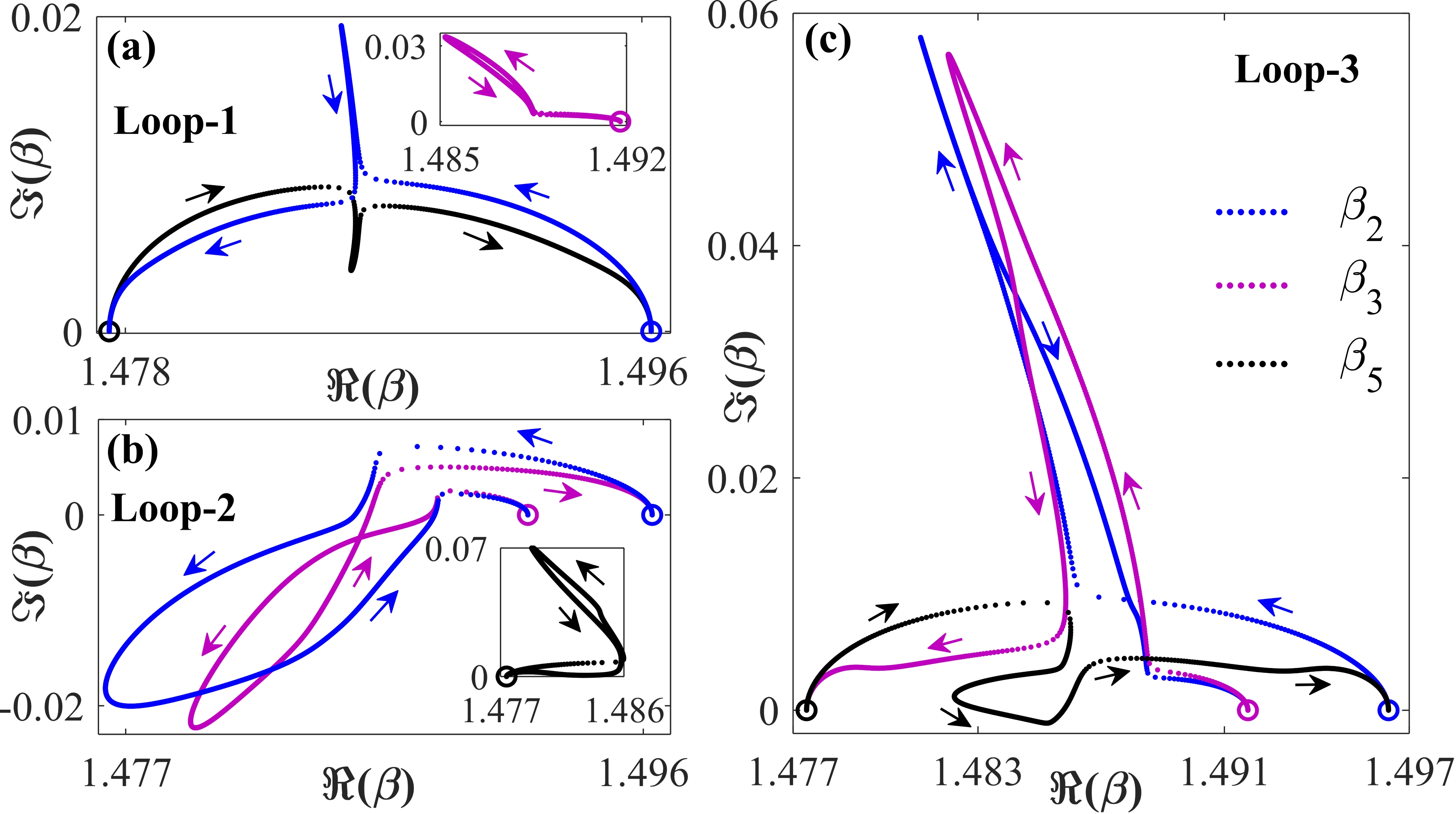}
\caption{Dynamics of complex $\beta_2$ (dotted blue trajectory ), $\beta_3$ (dotted pink trajectory) and $\beta_5$ (dotted black trajectory) following the adiabatic parametric variation governed by \textbf{(a)} Loop1 that encloses only EP2$^{(1)}$, \textbf{(b)} Loop2 that encloses only EP2$^{(2)}$, and \textbf{(c)} Loop3 that encloses both EP2$^{(1)}$ and EP2$^{(2)}$. The trajectories of $\beta_3$ in (a) and $\beta_5$ in (b) has been shown in the respective insets for clear visibility. Circular markers of the respective colours indicate the initial location of complex $\beta$-values.}
\label{fig_4}
\end{figure}
Here, $\gamma_0$, $\tau_0$, $a$ and $\phi\,(\in[0,2\pi])$ are the characteristics parameters. We have to vary $\phi$ very slowly for adiabatic approximation. To encircle an EP properly, we have to choose a $\gamma_0$ that must be greater than the $\gamma$-value of the respective EP. Now, we consider three different parametric loops to encircle both EP2s individually and also simultaneously, as shown in Fig. \ref{fig3} (a). Here, loop-1 (for $\gamma_0=0.008,\,\tau_0=4.5$ and $a=1.5$) encloses only EP2$^{(1)}$ and loop-2 (for $\gamma_0=0.02,\,\tau_0=4.5$ and $a=2$) encloses only EP2$^{(2)}$, whereas loop-3 (for $\gamma_0=0.015,\,\tau_0=4.8$ and $a=2.5$) encloses both the EP2s. Now, if we study the propagation of eigenmodes along the length of the waveguide following the encirclement processes, we have to consider a length-dependent loss-gain variation to encircle single or multiple EPs dynamically. Here, to consider dynamical encirclement process, we have to map the closed parametric loop given by Eq. \ref{device} throughout the length of the waveguide and that has been achieved with the consideration of $\phi=2\pi z/L$. Thus, for dynamical encirclement, Eq. \ref{device} can be rewritten as 
\begin{equation}
\gamma(z)=\gamma_{0}\sin\left[\frac{\pi z}{L}\right];\,\,\tau(z)=\tau_{EP}+a\sin\left[\frac{2\pi z}{L}\right].
\label{device1} 
\end{equation}
According to the defined shape of the parameter loop in $(\gamma, \tau)$ plane, $\gamma$ must be equal to zero for both input ($z=0$) and output ($z=L$). Thus, we can excite and retrieve the passive modes at input and output, respectively;  which is not  possible by using conventional circular parametric encirclement processes \cite{atomic_spectra}. The different directions of encirclement are realized by simply changing the direction of the propagation along the length of the waveguide. Following Eq. (\ref{device1}), the overall distribution of $\Im[n(x,z)]$ towards the length of the waveguide, correspond to the loop-3 have been shown in Fig. \ref{fig3}(b).

Now, we study the the effect of parametric encirclement on the propagation constants of three interacting modes $\beta_2,\,\beta_3$ and $\beta_5$ in the complex $\beta$-plane for each of the encirclement process in ($\gamma,\tau$)-plane shown in Fig. \ref{fig3}(a). We consider encirclement in clockwise direction with enough small steps (almost quasi statically), so that the adiabatic motion of corresponding modes are properly traced in Fig. \ref{fig_4}. Here, blue, pink, and black dotted curves represent the trajectories of the $\beta_2$, $\beta_3$ and $\beta_5$, respectively. Circular markers of the respective colours indicate the starting location of respective modes in complex $\beta$-plane. 

In Fig. \ref{fig_4}(a), we consider the encirclement process following loop-1 that encloses only EP2$^{(1)}$. Here, as $\beta_2$ and $\beta_5$ are analytically connected through EP2$^{(1)}$, they exchange their initial positions adiabatically and form a complete loop in complex $\beta$-plane for total $2\pi$ encirclement. However, this encirclement process does not effect the dynamics of $\beta_3$, which remain in the same state at the end of the encirclement process, as can be seen in the inset of Fig. \ref{fig_4}(a). Now, while we consider the parametric encirclement around EP2$^{(2)}$ following loop-2, we can observe the adiabatic exchange phenomenon between $\beta_2$ and $\beta_3$ in a generic fashion [as shown in Fig. \ref{fig_4}(b)]. Interestingly, in this case,  $\beta_5$ remains in the same state at the end of the encirclement process, that has been shown in the inset of Fig. \ref{fig_4}(b). Thus the individual encirclement around each of the EP2s in parameter plane allows the flipping between two corresponding coupled states in complex $\beta$-plane even in the presence of a other states, which solidly confirm their second-order branch point behavior of eigenvalues.
\begin{figure}[b!]
\centering
\includegraphics[width=12cm]{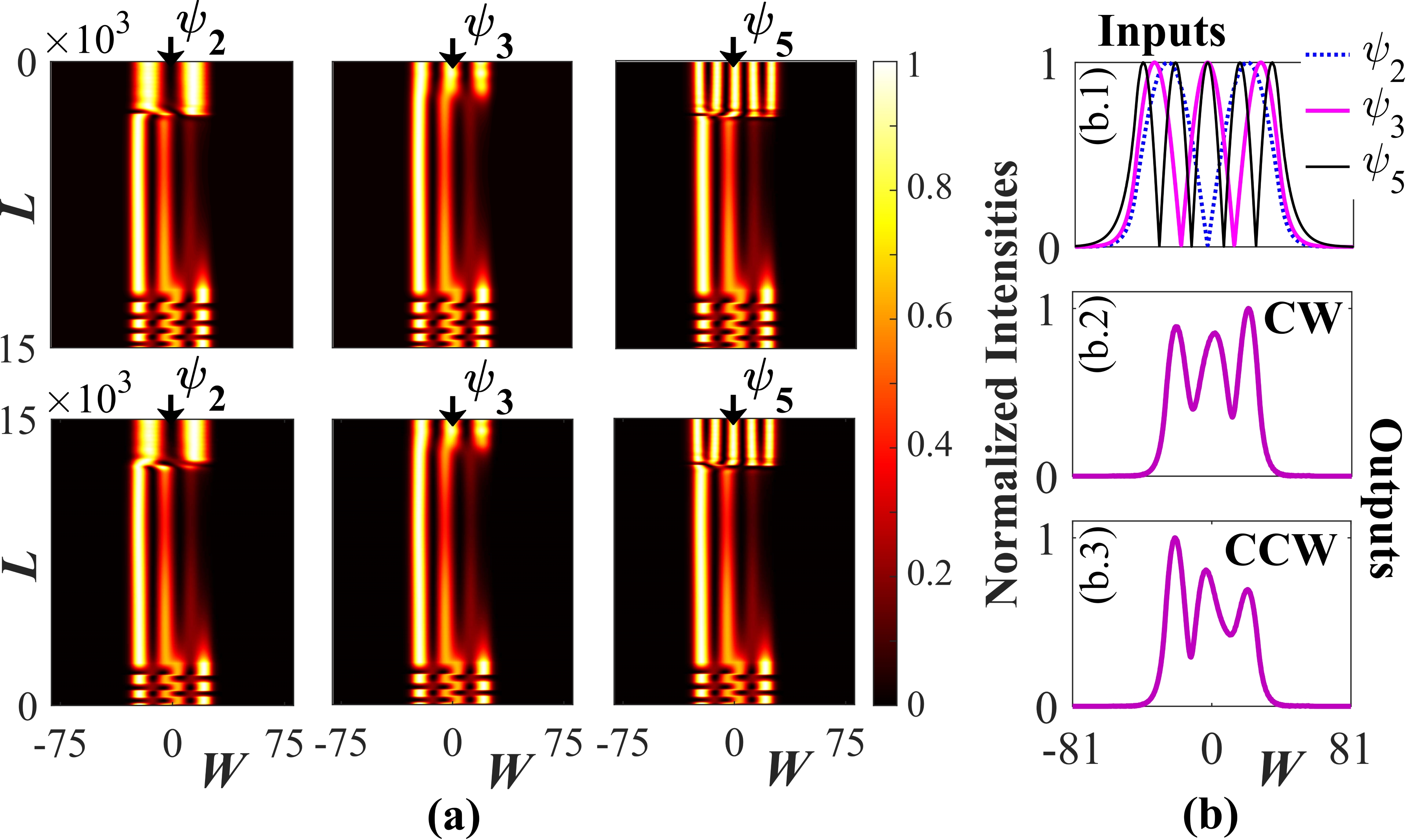} 
 \caption{\textbf{(a)} Beam propagation results of $\psi_j$ (j=2,3,5) following the dynamical encirclement scheme as shown in Fig. \ref{fig3}(b). The upper panel represent the beam propagation for encirclement in clockwise (CW) direction and the lower panel represent the beam propagation for encirclement in counter-clockwise (CCW) direction, where for both the cases, we find the conversions $\{\psi_2,\psi_3,\psi_5\}\rightarrow\psi_3$. We re-normalize the field intensities at each z for clear visualization. \textbf{(b)} (b.1) Normalized input field intensities; (b.2) normalized common field intensities at $z=L$ for CW encirclement and (b.3) at $z=0$ for CCW encirclement.}
\label{fig5}
\end{figure}
 
Now, if we enclose both EP2s in $(\gamma,\tau)$-plane by quasi-static parametric variation along the loop3, then the system shows third-order branch point behavior for the corresponding eigenvalues, and we trace the topological dynamics of three coupled modes in complex  $\beta$-plane. Corresponding results has been showing in Fig. \ref{fig_4}(c). Here, we have shown that the following one complete loop in $(\gamma,\tau)$-plane, all three coupled modes $\beta_2$, $\beta_3$ and $\beta_5$ are exchanging their own identities. They switch successively in a manner like $\beta_2$ $\rightarrow$ $\beta_3$ $\rightarrow$ $\beta_5$ $\rightarrow$ $\beta_2$ and make a complete loop in complex $\beta$-plane. This type of unconventional dynamics of three coupled modes around two EP2s indeed confirms the appearance of an EP3 in system parameter space, where all three chosen modes are analytically connected. The effect of parametric evolution, as shown in Fig. \ref{fig_4}(c), reveals the third-order branch point behavior due to presence of an EP3 \cite{atomic_spectra,EP3_sayan} because $\beta$-values of three coupled modes regain their initial locations after completing three consecutive parametric encirclement processes. In a similar way, with proper modulation of gain-loos profile, we can host multiple EP3s among possible sets of three interacting modes from six quasi-guided modes $\psi_j\,(j=1,2,3,4,5,6)$.

Now, we consider a dynamical encirclement scheme around the embedded EP3 by mapping the parametric variation governed by loop-3 in Fig. \ref{fig3} (that encloses both the EP2s) throughout the length of the waveguide following Eq. \ref{device1}. The corresponding dynamical parametric variation has already been shown in Fig. \ref{fig3}(b). Now, the beam propagation results of the chosen interacting modes $\psi_j\,(j=2,3,5)$, which are analytically connected at the EP3, have been shown in Fig. \ref{fig5}. In the upper panel of Fig. \ref{fig5}(a), we have considered the dynamical encirclement process in clockwise direction by choosing the input at $z=0$. Here, we observe that beyond the adiabatic expectation, all the three modes $\psi_2,\,\psi_3$ and $\psi_5$ are finally converted into $\psi_3$ at the output $(z=L)$ after completing the propagation in the forward direction. Thus, there are two nonadiabatic evolutions of $\psi_3$ that remains in $\psi_3$ and $\psi_5$ that is converted to $\psi_3$ [unlike the corresponding trajectories of $\beta_3$ and $\beta_5$ shown in Fig. \ref{fig_4}(c)]. Here, $\psi_2$ evolves adiabatically and converted into $\psi_3$ at the end of the encirclement. Now, if we consider dynamical encirclement process following the parametric variation shown in Fig. \ref{fig3}(b) in anticlockwise direction by exciting the modes form $z=L$, then also all the three modes are converted into $\psi_3$ at $z=0$ after completing the propagation in the backward direction. Corresponding results have been shown in the lower panel of Fig. \ref{fig5}(a). 

Thus, due to dynamical encirclement around both the connecting EP2s in any of the directions, we get a  fixed output solution $\psi_3$. This means that the device does not follow the chirality in presence of an EP3. Here, $\psi_2$ evolves with the lowest average loss among three interacting states that results in adiabatic evolution. A comparative study has been presented in Fig. \ref{fig5}(b), where we have shown the output field intensities for encirclement in both the directions along with the chosen input field intensities. Fig. \ref{fig5}(b.1) depicts the normalized field intensities at the input, whereas plots \ref{fig5}(b.2) and \ref{fig5}(b.3) represent the normalized output field intensities (like $\psi_3$) for clockwise (CW) and counter-clockwise (CCW) dynamical encirclement scheme, respectively. Now, we calculate the mode conversion efficiencies in term of the overlap integrals between the input output fields. The expression of conversion efficiency for a certain conversion $\psi_m$ $\rightarrow$ $\psi_n$ can be written as
\begin{equation}\label{conversion}
  C_{m\rightarrow n}=\frac{\left|\int\psi_m \psi_n dx\right| ^2} {\int\left|\psi_m\right| ^2 dx {\int\left|\psi_n\right| ^2 dx}};\quad(m,n) \in j,m\ne n
\end{equation}
Here, during clockwise evolution, we found maximum conversion efficiency of 78\%, whereas for anticlockwise dynamical encirclement, we achieve maximum 73\% conversion efficiency. 

In summary, we have reported the presence of an EP3 with simultaneous presence of two connecting EP2s in a six-mode supported planar waveguide geometry with a special type of customized gain-loss profile in terms of two tunable parameters. We have verified the third-order branch point behavior with a successive $\beta$-switching phenomenon between three chosen interacting modes following an adiabatic encirclement process around both the connecting EP2s. Here, we have shown the breakdown in chirality during dynamical parametric evolution in the vicinity of the embedded EP3, where all the three interacting modes are converted to a specific dominating mode irrespective of the direction of encirclement in the sense of the direction of propagation through the waveguide. The potential applicability of the proposed design through state-of-the-art fabrication techniques can be further explored as a higher-order mode converter that can excite a particular mode in a multi-mode configuration. The proposed scheme should facilitate the photonic devices on the chip-scale device footprint for the next-generation optical system.

\textbf{Funding.} Ministry of Human Resource Development (MHRD) and Science and Engineering Research Board [Grant No. ECR/2017/000491], Ministry of Science and Technology, India.

\nocite{*}

\bibliography{my_ref}

\end{document}